\documentstyle[aps,prd]{revtex}
\begin{document}

\newcommand{\be}{\begin{equation}}
\newcommand{\ee}{\end{equation}}
\newcommand{\ben}{\begin{eqnarray}}
\newcommand{\een}{\end{eqnarray}}
\newcommand{\n}{\label}
\newcommand{\no}{\noindent}
\def \l {\Lambda}
\def \o {\Omega}

\title{Enlarged quintessence cosmology}

\author{
Luis P. Chimento\footnote{Electronic address:
chimento@df.uba.ar}  and  Alejandro S. Jakubi\footnote{
Electronic address: jakubi@df.uba.ar}}
\address{Departamento de F\'{\i}sica, Universidad de 
Buenos Aires, 1428~Buenos Aires, Argentina}
\author{Diego Pav\'{o}n\footnote{Electronic address:
diego@ulises.uab.es}}
\address{
Departament de F\'{\i}sica, Universidad Aut\'onoma 
de Barcelona, 08193 Bellaterra, Spain}

\date{\today}

\maketitle


\begin{abstract}

We show that the combination of a fluid with a bulk dissipative pressure and
quintessence matter can simultaneously drive an accelerated expansion phase
and solve the coincidence problem of our current Universe. We then study some
scenarios compatible with the observed cosmic acceleration.
  
\end{abstract}

%==========================================================================
\section{Introduction}
%==========================================================================

In the wake of the recent measurements of distant exploding stars, 
supernovae, the existence of negative-pressure dark energy has
begun to gain broad consideration.  Using Type Ia supernovae as standard
candles to gauge the expansion of the Universe, observers  have found evidence
that the Universe is accelerating~\cite{Perlmutter98,Riess98,Garnavich98}. 
A new component with significant negative pressure, called quintessence matter
(Q-matter for short) will in fact cause the
cosmic expansion to speed up, so the supernovae observations provide 
empirical support for a new form of energy with strong negative
pressure~\cite{Perlmutter98,OS95,Peeb84,Wang99,CDS98,Perlmutter99,Scott}.

Different forms for the quintessence energy have been proposed. They include 
a cosmological constant (or more generaly a variable cosmological
term), a scalar field \cite{CDS98}, a frustrated network of non--Abelian 
cosmic strings, and a frustrated network of domain walls
\cite{Bucher98}, \cite{Bucher99}. All these proposals assume the 
Q-matter behaves as a perfect fluid with a linear baryotropic
equation of state, and so some effort has been invested in determinating 
its adiabatic index at the present epoch -see e.g. \cite{Wang99,Waga}. 
This new energy is to be added to the more familiar components: 
i.e., normal matter (luminous and dark) plus radiation. 
The contribution of the radiation component is known to be negligible 
at the current epoch whereas the main contribution to the former comes 
from cold dark matter (CDM) \cite{Peacock}.

However, QCDM models (including those in which the quintessence energy is just
the energy of the quantum vacuum) find dificulties in explaining why the
energy densities of the CDM and Q--matter should be comparable today. Since 
both energies redshift at different rates the conditions at the early 
universe have to be set very carefully for both  energy densities to be of the 
same order today, though one may always invoke some version or other  of the
anthropic principle to soften a bit the problem. This is the coincidence 
problem \cite{CMBdata}. Recently, a promising solution, for
spatially--flat metrics, based in the notion of ``tracker fields", fields
that roll down their potential according to an attractorlike solution to
the equations of motion \cite{zlatev,Wang99,Scott}, has been proposed.
Unfortunately it is not clear to what extent these fields have the ability 
to solve the coincidence problem and, at the same time, drive the Universe 
to the current phase of accelerated expansion.   

In any case, all these models overlook the fact that since the cosmic fluid
consists in a mixture of different fluids a dissipative pressure may naturally
arise which, for expanding universes, is bound to further decrease the total
pressure \cite{dw93}. Recently it has been proposed that the CDM must
self--interact to explain the structure of the halos of the galaxies
\cite{self} -see however \cite{Moore}. This self--interaction leads naturally
to a viscous pressure whose magnitud will depend on the mean free path of the
CDM particles. On the other hand it has been suggested that dissipative fluids
(or equivalently particle production processes) can drive a phase of
accelerated expansion -see \cite{Barrow} and references therein.

This paper investigates how the combined action of dissipative normal matter
and a quintessence scalar field may lead to the current accelerated expansion
stage, and at the same time  provide a solution to the coincidence problem
different from that relying on a tracker field. It is shown in Sec. II that
the flat coincident solution is an attractor. Section III deepens in the
consequences brought about by a dissipative pressure in the strees--energy
tensor, i.e., it explores the dynamics of quintessence--dissipative dark
matter (QDDM) models, while Sec. IV briefly outlines some specific models.
Finally, Sec. V summarizes our findings. It should be understood that this
work mainly deals with the present and late Universe since it is precisely at
these stages where the dynamic effects of quintessence and dissipative
pressure become important. Units in which $c = 8\pi G = k_{B} = 1$ are used
throughout.

%==========================================================================
\section{Cosmological problems}
%==========================================================================

This section shows that the Friedmann-Lamaitre-Robertson-Walker (FLRW)
universe filled with perfect normal matter plus quintessence fluid,
corresponding to some scalar field governed by Klein--Gordon equation, cannot
at the same time drive an accelerated expansion and solve the coincidence
problem. To solve it, without abandoning the FLRW geometry, some additional
contribution to the stress--energy tensor, such as a bulk dissipative
pressure, is needed.

The overall stress--energy tensor of the cosmic fluid without the 
dissipative pressure reads
\be
\label{set}
T_{ab} = \rho u_{a} u_{b} + p h_{ab}\, , \quad 
(h_{ab} = g_{ab} + u_{a} u_{b}\,  \quad u^{a} u_{a} = - 1),
\ee

\noindent where $\rho = \rho_{m} + \rho _{\phi}$  and $p = p_{m} + p_{\phi}$.
Here $\rho_m$ and $p_m$ are the energy density and pressure of the matter
whose equation of state is $p_{m} = (\gamma_{m} -1) \rho_{m}$ with adiabatic
index in the interval $ 1 \leq \gamma_{m} \le 2$. Likewise $\rho_{\phi}$ and
$p_{\phi}$, the energy density and pressure of the minimally coupled
self--interacting Q-matter field $\phi$, i.e.,

\be
\label{prhophi}
\rho_{\phi}  =   \frac{1}{2} \dot{\phi}^{2} + V(\phi)\, ,
\qquad
p_{\phi} =   \frac{1}{2} \dot{\phi}^{2} - V(\phi) \, , 
\ee
\noindent
are related by an equation of state similar to that of the matter,
{\it viz.} $p_{\phi} = (\gamma_{\phi} - 1) \rho_{\phi}$, so that its adiabatic
index is given by

\begin{equation} \label{gammaphi}
\gamma_\phi=\frac{\dot\phi^2}{\dot\phi^2/2+V(\phi)},
\end{equation}

\noindent 
where for non--negative potentials $V(\phi)$ one has $0 \leq
\gamma_{\phi} \leq 2$. The scalar field can be properly interpreted as
Q--matter provided $\gamma_{\phi} < 1$ -see e.g. \cite{zlatev}. As usual an
overdot means derivative with respect to cosmic time. In general
$\gamma_{\phi}$ varies as the Universe expands, and the same is true for
$\gamma_{m}$ since the massive and massless components of the matter fluid
redshift at different rates.

The Friedmann equation together with the energy conservation of the normal 
matter fluid and quintessence (Klein-Gordon equation) are

\be
\label{feq}
H^{2}  + \frac{k}{a^{2}} =  \frac{1}{3}( \rho_{m} + \rho_{\phi}) 
\qquad (k = 1, 0, -1),
\ee

\be
\dot{\rho_m}  +  3 H\gamma_m \rho_m  = 0 ,
\label{drho}
\ee

\be
\ddot{\phi}  +  3H \dot{\phi}  + V' = 0 ,
\label{KG}
\ee
\noindent
where $H \equiv \dot{a}/a$ denotes the Hubble factor and
the prime means derivative with respect to $\phi$.
\noindent
Introducing $\Omega_m \equiv \rho_{m} /\rho_{c}$, 
$\Omega_{\phi} ,\equiv \rho_{\phi} /\rho_{c}$, with
$\rho_{c} \equiv 3 H^{2}$ the critical density, and 
$\Omega_k \equiv-k/(aH)^2$ plus the definition 
$\Omega \equiv \Omega_{m} + \Omega_{\phi}$
the set of equations (\ref{feq}--\ref{KG}) 
can be recast as (cf. \cite{Ellis})

\be
\label{constr}
\Omega_{m} + \Omega_{\phi}+\Omega_k=1,
\ee

\be
\label{dOmega}
\dot{\Omega} = \Omega \left(\Omega - 1\right) \left(3 \gamma - 2\right) H \, ,
\ee

\be
\label{dOmegaphi}
\dot{\Omega}_{\phi} = \left[2 + \left(3 \gamma - 2\right) \Omega - 3 
\gamma_{\phi}\right] \Omega_{\phi} H \, ,
\ee

\noindent
where $\gamma$ is the average adiabatic index given by

\begin{equation} \label{gammaOmega}
\gamma\Omega=\gamma_m\Omega_m+\gamma_\phi\Omega_\phi.
\end{equation}

Next subsections investigate the flatness and coincidence problems. The 
former will be solved if the solution $\Omega=1$ to equation (\ref{dOmega}) 
becomes an attractor at late time. In its turn, the coincidence problem 
will be solved if the ratio $\Omega_{\phi}/\Omega_{m}$ becomes asymptotically 
a constant. We shall therefore  explore the possibility of constant stable 
solutions in the $(\Omega,\Omega_m,\Omega_\phi)$ space.

\subsection{The flatness problem and accelerated expansion}

The combined measurements of the cosmic microwave background  temperature
fluctuations and the distribution of galaxies on large scales seem to imply
that the Universe may  be flat or nearly flat \cite{OS95,Neta99,Efst}.
Hence the interesting solution at late times of (\ref{dOmega}) is $\Omega = 1$
(i.e., $ k= 0$), and so we discard the solution $\Omega=0$ as incompatible
with observation. The solution $\Omega=1$ is asymptotically stable
for  expanding universes ($H > 0$) provided that the condition
$\partial\dot\Omega/\partial\Omega<0$ holds in a neighborhood of $\Omega=1$
and this implies $\gamma < 2/3$. Hence the matter stress violates the
strong energy condition (SEC) $\rho +3p\ge 0$ and as a consequence
the Universe accelerates its expansion, i.e., 
$\ddot{a}/a = -(\rho+3p)/6 > 0$.

Let us examine more closely the implications of the current accelerated
expansion for the QCDM model. Since the mixture of Q--matter and
perfect dark matter fluid violates the SEC, $\gamma_{\phi}$ must be 
low enough. Namely, since $\gamma<2/3$, $\gamma_{m}\ge 1$, and
$\gamma_{\phi} < \gamma_{m}$, equation (\ref{gammaOmega}) implies 
$\gamma_{\phi} < \gamma$. Then, introducing $\Omega = 1$ in equation 
(\ref{dOmegaphi}) we obtain

\begin{equation} \label{dOmegaphi2}
\dot{\Omega}_{\phi} = 3 (\gamma - \gamma_{\phi}) \Omega_{\phi} H,
\end{equation}

\noindent 
and therefore $\dot{\Omega}_{\phi} > 0$, i.e., $\Omega_{\phi}$ will
grow until the constraint (\ref{constr}) is saturated, giving 
$\Omega_\phi= 1$ in the asymptotic regime. Thus the matter 
fluid yields a vanishing contribution to the energy density 
of the Universe at large time. This implies that a flat
FLRW universe driven by a mixture of normal perfect fluid and 
quintessence matter cannot both drive an accelerated expansion 
and solve  the coincidence problem. Therefore some other 
contribution must enter the stress--energy tensor of the 
cosmic fluid, i.e., it must be ``enlarged".

\subsection{The coincidence problem}

As shown above we cannot have both the current accelerated 
expansion and the coincidence problem solved within a model
that assumes a perfect matter fluid and a quintessential scalar field. 
Things fare differently when a dissipative pressure enters the play. 
Because of the FLRW metric, velocity gradients causing shear 
viscosity and temperature gradients leading to heat transport must be
absent. Therefore the only admissible dissipative term 
corresponds to a bulk dissipative pressure $\pi$. 
This quantity is always negative for expanding fluids (i.e., $\pi < 0$
so long  as $H > 0$) and may be understood either as a viscous pressure 
or as the effect of particle production. On general grounds the former
possibility is usually thought to give just a small contribution to the
overall pressure, however the impact of the latter is not so much limited. 
The expression of the bulk stress when interpreted that way is 
$ \pi = -(\rho_{m} + p_{m}) \Gamma/3 H$ where $ \Gamma$ denotes the 
particle production rate. This process is dissipative in the sense
that the produced particles imply an augment of the phase space volume.
A recent discussion about the interplay between dissipative bulk 
pressure and cosmological particle production can be found in 
\cite{zimdahl}.

So the total stress--energy tensor of the cosmic medium, made up of 
a dissipative but otherwise normal fluid plus the Q--matter fluid, 
reads

\begin{equation}
T_{ab} = (\rho_m+\rho_{\phi} + p_m+p_{\phi} +\pi) u_{a} u_{b} +
(  p_m+p_{\phi}+\pi) g_{ab}.
\label{2}
\end{equation}

\no 
A parallel calculation to that of above leads to
the corresponding Einstein-Klein-Gordon field equations

\be
\label{dOmegapi}
\dot{\Omega} = \Omega \left(\Omega - 1\right)
\left[3 \left(\gamma + \frac{\pi}{\rho}\right) - 2
\right]
H \, ,
\ee

\noindent
and

\be
\label{dOmegaphipi}
\dot{\Omega}_{\phi} = \left\{2 + \left[3 \left(\gamma +
\frac{\pi}{\rho}\right) - 2\right]
\Omega - 3 \gamma_{\phi}\right\} H \Omega_{\phi},
\ee
\noindent
instead of equations (\ref{dOmega}) and (\ref{dOmegaphi}). 
The energy conservation of the normal matter is

\be
\label{drhompi}
\dot{\rho_{m}} + 3\left(\gamma_{m}+ 
\frac{\pi}{\rho_{m}}\right) \rho_{m} H = 0.
\ee

\noindent
Owing to the presence of the dissipative bulk stress the 
constraint $ \gamma < 2/3 $ does not longer have to be fulfilled 
for the solution $\Omega = 1$ of equation (\ref{dOmegapi}) to be
stable. Likewise, inspection of (\ref{dOmegaphipi}) shows that 
when $\Omega = 1$  one can have $\dot{\Omega}_{\phi} <0 $ just 
by choosing the ratio $\pi/\rho$ sufficiently negative . 
Thereby the constraint (\ref{constr}) allows a nonvanishing
$\Omega_{m}$ at large times. By contrast tracker fields based models
(valid only when $\Omega_{k} = 0$) predict that 
$\Omega_{m} \rightarrow 0$ asymptotically \cite{zlatev}. 
\noindent
A fixed point solution of equation (\ref{dOmegapi}) is $\Omega=1$. Note
that equations (\ref{constr}) and (\ref{dOmegaphipi}) have fixed point 
solutions $\Omega_m=\Omega_{m0}$ and $\Omega_\phi=\Omega_{\phi 0}$,
respectively, when the partial adiabatic indices and the dissipative 
pressure are related by

\begin{equation} \label{gammapi}
\gamma_{m}+\frac{\pi}{\rho_{m}}=\gamma_{\phi} = -\frac{2\dot{H}}{3H^2}.
\end{equation}

\noindent
Then smaller $\gamma_\phi$, the larger the dissipative effects.
Let us investigate the requirements imposed by the stability of these
solutions. From (\ref{dOmegapi}) we see that $\gamma+\pi/\rho<2/3$ must be
fulfilled if the solution $\Omega=1$ is to be asymptotically stable. This
condition, together with (\ref{gammapi}), leads to the additional constraint
on the viscosity pressure

\begin{equation} \label{pic}
\pi<\left(\frac{2}{3}-\gamma_m\right)\rho_{m} \, ,
\end{equation}

\noindent 
which is negative for ordinary matter fluids. Also by virtue of
(\ref{dOmegaphi2}) and the first equality in (\ref{gammapi}) we obtain 
from (\ref{pic}) that $\gamma_{\phi}<2/3$.

In the special case of a spatially flat universe ($\Omega=1$), the
stability of the solutions $\Omega_{m0}$ and $\Omega_{\phi 0}$ may be studied
directly from (\ref{dOmegaphipi}). Namely, setting 
$\Omega_\phi=\Omega_{\phi0}+\omega$ and using 
(\ref{gammaOmega}) it follows that

\begin{equation} \label{}
\dot\omega=3\Omega_{m} \left(\gamma_m-\gamma_\phi+
\frac{\pi}{\rho_{m}}\right) H \left(\Omega_{\phi 0}+\omega\right).
\end{equation}

\noindent
Accordingly the solution $\Omega=1$, $\Omega_\phi=\Omega_{\phi 0}$ is stable 
for the class of models that satisfies 
$\psi \equiv\gamma_m-\gamma_\phi+(\pi/\rho_m)<0$
and $\psi\to 0$ for $t\to\infty$. Note that this coincides 
with the attractor condition (\ref{gammapi}).

In order to study the stability of the solutions $\Omega_{m0}$ and
$\Omega_{\phi 0}$ when $k\neq 0$ it is advisable to derive a dynamic 
equation for the density ratio parameter

\[ \epsilon \equiv \frac{\Omega_{m}}{\Omega_{\phi}}. \]

\noindent For this purpose we combine the
logarithmic derivative of $\epsilon$ with the definitions of $\Omega_{m}$ and
$\Omega_{\phi}$ and the energy conservation equations (\ref{drhompi}) and
(\ref{KG})  -the latter written in terms of $\rho_{\phi}$. It yields

\begin{equation} \label{depsilon}
\dot{\epsilon} = 3\left(\gamma_{\phi}-\gamma_{m}-
\frac{\pi}{\rho_{m}}\right)H\epsilon.
\end{equation}

\noindent 
To calculate $\gamma_{\phi}$ we use (\ref{drhompi})
together with (\ref{gammaphi}), (\ref{feq}) and (\ref{KG}), obtaining

\begin{equation} \label{gammaphi0}
\gamma_\phi=\gamma_m+\frac{\pi}{\rho_m}
-\frac{1}{\Omega_\phi}\left[\frac{2\dot H}{3H^2}+
\gamma_m+\frac{\pi}{\rho_m}+
\left(\frac{2}{3}-\gamma_m-\frac{\pi}{\rho_m}\right)\Omega_k\right].
\end{equation}

\noindent
Introducing (\ref{gammaphi0}) in (\ref{depsilon}) we get

\begin{equation} \label{depsilon2}
\dot\epsilon=
-\frac{3H\epsilon}{\Omega_\phi}\left[\frac{2\dot H}{3H^2}+
\gamma_m+\frac{\pi}{\rho_m}+
\left(\frac{2}{3}-\gamma_m-\frac{\pi}{\rho_m}\right)\Omega_k\right],
\end{equation}

\noindent
and perturbating this expression about the solution 
$\epsilon_{0} \sim {\cal O}(1)$, 
(i.e., using the ansatz $\epsilon=\epsilon_{0}+\delta$ with $|\delta|\ll 1$) 
we obtain with the help of (\ref{gammapi})

\begin{equation}
\n{delta2}
\dot\delta=-\frac{3}{\Omega_\phi}\left(\frac{2}{3}-
\gamma_{\phi}\right)\Omega_{k} H \left(\epsilon_0+\delta\right)
\end{equation}

\noindent
near the attractor. For $\Omega_{k} > 0$ (negatively spatially curved
universes) it follows that $\delta$ decreases, i.e., the ratio
$(\Omega_{m}/\Omega_{\phi})_{0}$ is a stable solution. For $\Omega_{k} < 0$
one has to go beyond the linear perturbative regime and/or restrict 
the class of models as in the spatially flat case to determine the
stability of the solution. We defer this to a future research.

As we mentioned above, recently there have been some claims that CDM must not
be a perfect fluid because it ought to self--interact (with a mean free path
in the range $1 \, \mbox{kpc} \leq l \leq 1 \, \mbox{ Mpc}$) if one wish to 
explain the structure of the halos of galaxies \cite{self}. In this light it 
is not unreasonable to think that this same interaction is the origin of the 
dissipative pressure $\pi$ at cosmological scales. Bearing in mind that  
$l = 1/n \sigma $, with $n$ the number density of CDM particles and 
$\sigma$ the interaction cross section, a simple estimation reveals that 
at such scales $l$ is lower than the Hubble distance $H^{-1}$ and 
accordingly  the fluid approximation we are using is valid.

%======================================================================
\section{QDDM asymptotic era}
%======================================================================

Bulk viscosity arises typically in mixtures--either of different particles 
species, as in a radiative fluid, or of the same species but with different 
energies, as in a Maxwell--Boltzmann gas. Physically, we can think of $\pi$ 
as the internal ``friction" that sets in due to the different cooling rates 
in the expanding mixture.

Any dissipation in exact FLRW universes have to be scalar in nature, and in 
principle it may be modelled as a bulk viscosity effect within a nonequilibrium
thermodynamic theory such as the Israel--Stewart's \cite{d1}, \cite{m2}. In 
that formulation, the transport equation for the bulk viscous pressure takes 
the form

\begin{equation}
\pi + \tau\dot{\pi}  =  - 3\zeta H -
\frac{1}{2}\pi \tau\left[3H + \frac{\dot{\tau}}{\tau}
- \frac{\dot{\zeta}}{\zeta} - \frac{\dot{T}}{T} 
\right],
\label{dpi}
\end{equation}

\noindent
where the positive--definite quantity $\zeta $ stands for the phenomenological 
coefficient of bulk viscosity, $T$  the temperature of the fluid,
and $\tau$ the relaxation time associated to the dissipative pressure
-i.e., the time the system would take to reach the thermodynamic 
equilibrium state if the velocity divergence were suddenly turned 
off \cite{mdv}. Usually $\zeta$ is given by the kinetic theory of 
gases or a fluctuation-dissipation theorem or both \cite{jpa}.

Provided the factor within the square bracket in (\ref{dpi}) is small
it can be approximated by the more manageable truncated transport equation

\begin{equation}
\pi + \tau\dot{\pi}  =  - 3\zeta H \, ,
\label{dpi2}
\end{equation}

\no
widely used in the literature. This as well as (\ref{dpi})
meets the requirements of causality and stability to be
fulfilled by any physically acceptable transport equation \cite{HL}.

%=======================================================================
\subsection{The quasiperfect regime}
%=======================================================================

Here we obtain an explicit expression for the leading behavior
of the attractor solution at late time. We begin by writing the
equation of motion for the Hubble factor that follows from combining  
(\ref{dOmegapi}) and (\ref{drhompi}) with (\ref{dpi2})

$$
\ddot{H}+3\gamma H\dot{H}+\tau^{-1}\left[\dot{H}+{\textstyle{3\over
2}}\left(\gamma+\tau\dot\gamma\right)H^{2} -
{\textstyle{3\over 2}}\,\zeta H\right]
$$
\begin{equation}
+\frac{k}{a^{2}} \left[(1- {\textstyle{3\over
2}}\gamma)\left(2H-\tau^{-1}\right)
+ {\textstyle{3\over 2}} \dot{\gamma}\right]=0.
\label{5}
\end{equation}

\noindent
We next evaluate (\ref{gammaOmega}) on the attractor and
insert it together with (\ref{gammapi}) in (\ref{5}) to
obtain

\begin{equation} \label{dHa}
\nu^{-1}\left(\frac{\ddot H}{H}+3\gamma_m \dot H\right)+
\dot H+\frac{3\gamma_m}{2}H^2
-\frac{3\zeta}{2\Omega_{m0}}H=0 \,.
\end{equation}

Observation seems to rule out huge entropy production processes on
large scales, otherwise the flux of gamma-rays we witness should 
be much higher \cite{Stecker}. Hence we shall assume that the 
viscous effects are not as large as that, but however not altogether 
negligible. If $\tau$ is the relaxation time, then 
$\nu=\left(\tau H\right)^{-1}$ is the number of relaxation times in 
a Hubble time -for quasistatic expansions $\nu$ is proportional to the 
number of particle interactions in a Hubble time. Perfect fluid behavior 
occurs in the limit $\nu\to\infty$, and a consistent hydrodynamical 
description of the fluids requires $\nu>1$. Thus we are lead to assume 
that $\tau H$ is small and we propose a ``quasiperfect" 
expansion in powers of $\nu^{-1}$.

Let us show that the attractor solution of leading behavior 
$a \simeq t^{\sigma}$ when $t\to\infty$, with $\sigma$ a 
positive--definite constant, is consistent in the quasiperfect regime. 
Indeed, by virtue of (\ref{gammapi}) it implies $\gamma_\phi\simeq 2/3\sigma$ 
and

\begin{equation} \label{pirhoa}
\frac{\pi}{\rho_m}\simeq
-\left(\gamma_m-\frac{2}{3\sigma}\right).
\end{equation}

\noindent
For approximately constant $\gamma_m$ ($\gamma_m=1$ for CDM), we get from
(\ref{pirhoa})

\begin{equation} \label{dpi0}
\frac{\dot\pi}{\pi}\simeq \frac{\dot\rho_m}{\rho_m}
\simeq -2\frac{H}{\sigma}.
\end{equation}

\noindent
Hence (\ref{dpi2}) becomes

\begin{equation} \label{pia}
\pi\left(1-\frac{2}{\nu\sigma}\right)\simeq -3\zeta H\, ,
\end{equation}

\noindent
and we get to leading order in $\nu^{-1}$

\begin{equation} \label{zetaa}
\zeta\simeq \Omega_{m0}\left(\gamma_m-\frac{2}{3\sigma}\right)H \, .
\end{equation}

\noindent Then integration of (\ref{drhompi}) yields $\rho_m\simeq
a^{-2/\sigma}$, the same scaling law as $\rho_\phi$. Finally its insertion
in (\ref{feq}) leads back to $a \simeq t^\sigma$, showing the consistency
of our assumptions. These results correspond to the lowest order 
in the quasiperfect expansion.

To go a step further we introduce the expansion of $H$ in powers of 
$\nu^{-1}$

\begin{equation} \label{H..}
H=H_0\left(1+h_1\nu^{-1} +\cdots\right)
\end{equation}
\noindent 
in (\ref{dHa}), and assuming that
$|\dot\tau|\ll \nu^{-1}$, it follows the approximated solution

\begin{equation} \label{Ha}
H\simeq\frac{\sigma}{t}\left[1+\left(\frac{3\gamma_m\sigma-2}{\sigma}+
\frac{\theta}{t}\right)\frac{1}{\nu} \right],
\end{equation}

\noindent
where $\theta$ is an arbitrary integration constant. This expression
reveals  that the power law is an attractor solution and that for $\sigma>
1$ (deceleration parameter $q = -(\sigma - 1)/\sigma<0$), CDM viscosity
provides an accelerated expansion scenario that also solves the
coincidence problem. It can be shown that $\gamma_\phi$ does not pick any
correction of order $t^{-1}$ from the subdominant term in (\ref{Ha}). Instead
the first correction appears to the order $\nu^{-2}t^{-2}$, and this fact
shows the high degree of correction of the approximation that 
$\gamma_\phi$ takes a constant value in the late time regime.

We note that this attractor solution works for any viscosity coefficient 
with leading behavior (\ref{zetaa}). In particular the case
$\zeta\propto\sqrt{\rho_m}$, investigated in \cite{Chi93,ChJMM}, 
satisfies this requirement.

%=======================================================================
\subsection{Full causal corrections}
%=======================================================================
 
Here we gauge the changes bring about by the the full transport equation 
(\ref{dpi}) on the expansion exponent obtained in the previous section. 
Using that equation and the viscosity coefficient found in (\ref{zetaa}) 
we get 

\begin{equation} \label{dHca}
\nu^{-1}\left\{\frac{\ddot H}{H}-\frac{1+2r}{2}\frac{\dot H^2}{H^2}+
\frac{3}{2}\left[\gamma_m\left(\frac{3}{2}-r\right)+1\right] \dot H+
\frac{9}{4}\gamma_m H^3\right\}+
\dot H+\frac{1}{\sigma}H^2=0 ,
\end{equation}

\noindent 
where, to estimate the corrections, we have assumed
that in the asymptotic regime $T \propto \rho^r$, with $r$
a positive--definite constant, and we have used that 
$\rho_m\simeq \epsilon_0\rho/\left(1+\epsilon_0\right)$ 
in this regime. This power law relationship is the simplest way to guarantee 
a positive heat capacity. Usually $p$, $\rho$, $T$ and the particle number 
density $n$ are equilibrium magnitudes related by equations
of state of the form $\rho=\rho(T,n)$ and $p=p(T,n)$. Further the
thermodynamic relation  

\begin{equation} \label{6.1}
\left(\frac{\partial\rho}{\partial n}\right)_{T}=\frac{\rho+p}{n}-
\frac{T}{n}\left(\frac{\partial p}{\partial T}\right)_{n}
\end{equation}
\noindent
holds. This directly follows from the requirement that the entropy is a 
state function \cite{textbook}. In the particular case of a material 
fluid with $\rho=\rho(T)$ and constant adiabatic index, this relation 
imposes the constraint $r=\left(\gamma-1\right)/\gamma$, so that 
$0\le r\le 1/2$ for $1\le\gamma\le 2$. Inserting (\ref{H..}) in 
(\ref{dHca}), and assuming that $|\dot\tau|\ll
\nu^{-1}$, we obtain the approximate solution

\begin{equation} \label{Hca}
H\simeq\frac{\sigma}{t}\left\{1+\left[-\frac{2}{\sigma}
+\frac{3}{2}\left(\gamma_m\left(\frac{1}{2}-r-\frac{3}{2}\sigma\right)
+1+\frac{2r+1}{3\sigma}\right)+
\frac{C}{t}\right]\frac{1}{\nu} \right\}.
\end{equation}

\noindent 
Comparison with (\ref{Ha}) shows that except when $\sigma\simeq 1$, the 
use of the complete
transport equation leads a to slighty slower rate of expansion at late
time. Also, in this regime, the equilibrium temperature decreases as
$T\sim t^{-2r}\sim a^{-2r/\sigma}$.

%===================================================================
\subsection{Late Q-matter dynamics }
%===================================================================

In virtue of (\ref{gammaOmega}) the density parameter ratio can be
written in terms of the adiabatic indices

\begin{equation} \label{epsilongamma}
\epsilon=\frac{\gamma-\gamma_\phi}{\gamma_m-\gamma}.
\end{equation}

\noindent 
Since $\gamma_m$ is approximately constant, if $\gamma\to\gamma_0$ 
in the asymptotic regime when $\epsilon\to\epsilon_{0}$,
then $\gamma_{\phi}$ must also approach a constant value. Hence 
from (\ref{gammaphi}) get the constraint

\begin{equation} \label{constrainta}
V(\phi)/\dot\phi^2\simeq C
\end{equation}

\noindent
with $C>1$ as $\gamma_\phi<2/3$. Potentials that satisfy this constraint 
have been investigated in \cite{Bar93} and \cite{Chi96}. Then 
(\ref{KG}) becomes

\begin{equation} \label{KGa}
\frac{\ddot\phi}{\dot\phi}+\frac{3H}{1+2C}\simeq 0 \, .
\end{equation}

An interesting potential that meets this constraint is 
the exponential potential

\begin{equation}
V(\phi)=V_{0}\exp(-A\phi),
\label{expp}
\end{equation} 
\noindent
where $A$ and $V_{0}$ are constants. Now,
integrating (\ref{KGa}) and using (\ref{constrainta}) we get

\begin{equation} \label{phitea}
\phi(t)\simeq \frac{1}{A}\left[\ln\frac{V_0A^2\gamma_\phi}
{2\left(2-\gamma_\phi\right)}+2\ln t\right].
\end{equation}

\noindent 
Hence $\phi$ slowly rolls down the exponential potential as
$\dot{\phi}\propto 1/t$ when $t\to\infty$. Also we find that 
$C \simeq (3\sigma-1)/2$ with $\gamma_\phi\simeq 2/3\sigma <2/3$ 
on the attractor, irrespective of $V_{0}$ and $A$.

Perfect fluid QCDM models based on the exponential
potential are ruled out by observations \cite{Wang99}. However we
shall demonstrate in the next secion that in the realm of QDDM models 
the exponential potential yields satisfactory results without any fine
tuning of the parameters.

%===========================================================================
\section{QDDM models}
%===========================================================================

This section explores the dynamical evolution  of a universe filled with a
viscous material fluid and a quintessence scalar field by resorting to models
based on simple relationships for the nonequilibrium quantities. This allows
us to explore the large dissipative regime where the nonequilibrium pressure 
has a magnitude comparable with the energy density. Recently 
tracker--field models with inverse power potentials have attracted much 
interest \cite{zlatev,RatraPeebles}. Here we will investigate some QDDM 
models with exponential potentials (\ref{expp}) such that for a wide
range of initial conditions the scalar field settles into an attractor
solution that depends only upon a few nonequilibrium thermodynamical
parameters, addressing the coincidence problem.

%=======================================================================
\subsection{Linear dissipative regime}
%=======================================================================

The linear regime  $\zeta = \alpha  H$, with $\alpha $ a constant
in the interval $0 < \alpha <1$, arises for instance when the coefficient 
of bulk viscosity takes the form of a radiating fluid. We further assume that
the number of interactions of a generic CDM particle in a Hubble time is 
larger than unity so that the hydrodynamic regime is respected. Here we 
investigate models with the limiting behavior $\gamma\to 2/3$ in the 
asymptotic regime. We begin by inserting the ansatz 

\begin{equation} \label{gamma4}
\gamma=\frac{2}{3}\left(1+\chi\right)
\end{equation}

\noindent 
in equation (\ref{5}). It is immediately seen that the latter splits in
two equations, namely

\begin{equation} \label{dH0}
\ddot H+(2+\nu)H\dot H+\nu\left(1-\frac{3}{2}\alpha\right)H^3=0 \, ,
\end{equation}
\noindent
and
\be
\n{chi}
\left[H^2+\frac{k}{a^2}\right]\dot\chi+\left[2H\left(\dot H-
\frac{k}{a^2}\right)+\tau^{-1}\left(H^2+
\frac{k}{a^2}\right)\right]\chi=0 \, .
\ee

\noindent
Replacing the solution of (\ref{chi}) in (\ref{gamma4}) it
follows,

\begin{equation} \label{gamma5}
\gamma=\frac{2}{3}\left(1+b\frac{a^{2-\nu}}{k+\dot a^2}\right) \, ,
\end{equation}

\noindent where $b$ is an arbitrary integration constant. This expression for
$\gamma$ will be sensible only if it meets the restriction $0\le\gamma\le 2$.
Then (\ref{dH0}) can be transformed into a linear differential equation
of second order whose general solution in parametrized form is already known
\cite{Chi93}, \cite{Guanajuato94}. In particular $a \propto t^\sigma$, 
where $\sigma$ is the largest root of 
$\nu(1-3\alpha/2)\sigma^2-(2+\nu)\sigma+2=0$,
is an asymptotic stable solution in the limit $t\to\infty$. Then, using
(\ref{gammaOmega}) together with the attractor conditions $\Omega=1$,
$\gamma=2/3$ and (\ref{gammapi}), we find

\begin{equation} \label{sigma}
\sigma=\frac{2\left(1-\Omega_{m0}\right)}{2-3\Omega_{m0}\gamma_m}.
\end{equation}

\noindent
Hence the quintessence adiabatic index $\gamma_{\phi} = 2/3\sigma$ depends 
solely on the dark matter parameters $\Omega_{m}$ and $\gamma_{m}$ in the 
asymptotic regime. Moreover, a relationship between the dissipative 
parameters $\alpha$ and $\nu$ follows from (\ref{pia}) and (\ref{gammapi}),
namely

\begin{equation} \label{alpha}
\alpha=\Omega_{m0}\left(1-\frac{2}{\nu\sigma}\right)
\left(\gamma_m-\frac{2}{3\sigma}\right),
\end{equation}

\noindent 
and the requirement $\alpha>0$ implies $\sigma\nu>2$ or
$\nu>\nu_{min}=3\gamma_\phi$. The same condition arises from the requirement
that $\gamma\to 2/3$ when $t\to\infty$. Equations (\ref{sigma}) and 
(\ref{alpha}) show that $\alpha$ grows with $\nu$ attaining
$\alpha_{max}=(3\gamma_m-2)\Omega_m/3(1-\Omega_m)$ in the limit 
$\nu\to\infty$.

As it follows from (\ref{sigma}) there will be  accelerated expansion 
(i.e., $\sigma > 1$) if $\Omega_{m0} < 2/3\gamma_m<2/3$ and  accordingly
we obtain a family of exact solutions describing a QDDM scenario that 
solves the coincidence problem regardless of the value of the spatial 
curvature. Figure 1 depicts the dependence of $\gamma_{\phi}$ on $\Omega_{m}$
when $\gamma_{m}=1$. To make a rough estimate of the cosmological parameters 
in the late time era we assume that our Universe is currently close to the 
asymptotic  attractor regime and use the current observational bounds. After 
\cite{Wang99} the combination of low redshift, type Ia supernovae and COBE 
measurements determines (for a spatially flat universe) the range 
$\Omega_m\sim 0.3-0.4$ and $\gamma_\phi<0.6$. From figure 1 it is
seen that our linear dissipative model satisfies comfortably these 
constraints. For $\Omega_{m} = 0.3$ we get from (\ref{sigma}) 
$\sigma\simeq 1.27$. This is fully consistent with current
estimations of $Ht$ today \cite{powerlaw}, as they provide a
lower bound for $\sigma$ in a universe that started only recently 
a phase of accelerated expansion and approaches asymptotically the 
attractor regime.

%===========================================================================
\subsection{Viscous speed regime}
%===========================================================================

This scenario is somewhat more general than the previous one. It arises
when the bulk viscosity coefficient is given in terms of the speed
of the bulk viscous signal $v$ by \cite{m2}

\begin{equation}
\frac{\zeta}{\tau} = v^{2} \gamma_m \rho_m.
\label{zeta}
\end{equation}

\noindent 
We further assume $\tau$ related to $H$ by the same expression as before,
only that to simplify the calculations we now take $\nu$ constant.
Hence (\ref{5}) becomes

\begin{equation} \label{dh}
\nu^{-1}\left[h''+3\gamma h'+3\gamma 'h-
9\frac{v^2\gamma_m\epsilon}{1+\epsilon} h\right]+h'+3\gamma h=0,
\end{equation}

\noindent
where
\begin{equation} \label{h}
h \equiv H^{2}+\frac{k}{a^{2}} \, ,
\end{equation}

\noindent 
and the prime indicates derivative with respect to $\eta \equiv \ln a$.
Here we have used the scale factor $a(t)$ as a coordinate instead of 
the cosmological time $t$,  i.e., $a$ is assumed to be a monotonic 
function of $t$.

Equation (\ref{dh}) is useful to study the asymptotic stability of FLRW
expansions at late time because it can be rewritten in terms of the 
derivative of a Lyapunov function \cite{Cesari}

$$
\frac{d}{d\eta}\left\{\frac{1}{2}h'^2+
\frac{3}{2}\left[\gamma '+\gamma\nu-3\frac{v^2\gamma_m\epsilon}{1+\epsilon}
\right]h^2\right\}=
-\left(3\gamma+\nu\right)h'^2
$$
\begin{equation} \label{dh2}
+\frac{3}{2}\left[\gamma '+\gamma\nu-
3\frac{v^2\gamma_m\epsilon}{1+\epsilon}\right]' h^{2}.
\end{equation}

\noindent 
If the adiabatic index does not decreases too fast in the attractor era, 
a sufficient condition for the Lyapunov function to have a minimum at
the phase space point $(h,h')=(0,0)$ is that $\nu>3v^{2}$. Within this 
scenario the parameters of the matter fluid can be taken as
quasistatic, yielding an asymptotically stable minimum. The leading 
behavior of the solutions in this quasistatic regime is given by

\begin{equation} \label{da}
h^{2} = c_{1} a^{\lambda_{1}}+c_{2} a^{\lambda_2},
\end{equation}

\noindent where

\begin{equation} \label{lambda}
\lambda_{1,2}=\frac{1}{2}\left\{-\left(3\gamma+\nu \right)
\pm\left[\left(3\gamma-\nu \right)^2+
36\gamma_{m} v^2\Omega_{m} \right]^{1/2}\right\}
\end{equation}

\noindent
and the parameters are evaluated at the asymptotic attractor era.
For large scale factor, equation (\ref{da}) reduces to
$h^{2} \simeq c_{1} a^{\lambda_{1}}$
when $-2<\lambda_1<0$, and we have once again a power-law
accelerated cosmic expansion with $\sigma=-2/\lambda_{1}$. 
Likewise equation (\ref{da}) reduces to $h^{2}\simeq c_{1} a^{-2}$ for
$\lambda_1=-2$, and  $h ^{2}\simeq 0$ for $\lambda_{1}<-2$. These 
two latter cases correspond to linear evolutions at late time.

Combining (\ref{pia}) and (\ref{gammapi}) we get

\begin{equation} \label{plambda}
\lambda^2+\left(3\gamma_m+\nu\right)\lambda+3\gamma_m(\nu-3v^2)=0
\end{equation}

\noindent and using (\ref{gammaOmega}) together with the attractor constraints
$\Omega=1$ and (\ref{gammapi}) we find that $\lambda_1=-3\gamma_\phi$ also
satisfies (\ref{plambda}). Hence we obtain

\begin{equation} \label{gammaphi2}
\gamma_\phi=\frac{1}{6}\left\{3\gamma_m+\nu
-\left[\left(3\gamma_m-\nu \right)^2+36\gamma_m v^2\right]^{1/2}\right\}.
\end{equation}

\noindent 
In this model the quintessence adiabatic index does depend
on the parameters $\nu$ and $v$ while it is independent of
the density parameter $\Omega_m$. We plot $\gamma_\phi$ in figure 2 
for $\gamma_m=1$. As it can be seen a wide range of the parameter 
space $(\nu, v)$ is consistent with a spatially--flat accelerated  
universe and such that $\gamma_\phi<0.6$ for any value of $\Omega_{m}$. 
This shows another solution to the coincidence problem for any value of 
the spatial curvature and compatible with accelerated expansion. It is 
also seen that dissipative effects enlarge the parameter space where 
observational data has to be fitted, but global dynamic information 
alone cannot determine the specific values of $\nu$ and $v$. Note that the 
smaller $\gamma_\phi$, the larger the dissipative contribution to the
sound speed, and the smaller the interaction rate.

%============================================================================
\section{Discussion}
%============================================================================

We have proved that the coincidence
problem and an accelerated expansion phase of FLRW cosmologies
cannot be simultaneously addressed by the combined effect of a
perfect fluid and Q-matter. Nonetheless, if nonbaryonic dark matter
behaves as a dissipative fluid rather than a perfect one, both problems may
find a simultaneous solution. This is so because an imperfect (i.e., 
dissipative fluid) expanding in a FLRW background possess a negative pressure 
$\pi$ that enters the conservation equations of general relativity. 
The models presented here are compatible with a negative deceleration 
parameter at present time. In consequence, the quintessence scenario 
becomes more robust when the dissipative effect of the nonequilibrium 
pressure arising in the CDM gas is allowed into the picture.
 
Recently attempts have been made to constraint the state equation of the
cosmic fluids (Q--matter included) by considering gravitational lensing
effects, the mass power spectrum and the anisotropies of the cosmic 
backgroud radiation \cite{Wang99},\cite{CHSN}. We have shown 
specific models with an ample region in the space of
out--of--equilibrium thermodynamic parameters satisfying this constraint in
the asymptotic attractor regime which our Universe may well be
approaching. We would like to point out that the parameter space should be
enlarged by adding these out--of--equilibrium parameters when fitting the 
observational data. Unfortunately there is some degenerancy in the 
determination of these parameters from constraints arising from the 
cosmological dynamics alone. We hope, however, that simulations of structure 
formation that include dissipative effects will ultimately prove 
instrumental in discriminating between different models.

On the attractor asymptotic regime the dissipative matter fluid and the 
scalar field contribute in a fixed ratio to the pressure and energy density 
along the QDDM era. This scenario ameliorates the self--adjusting
model \cite{Ferreira} as it allows for $0<\gamma<2$ for a wide range of 
initial conditions. It also improves on the tracking models as it solves the 
coincidence problem in the late accelerated expansion phase. While keeping 
a finite difference between the adiabatic indices of quintessence and 
matter fluid, this difference arises in the viscous pressure.

As it has been noted the Q--matter proposal may entail some undesirable
effects such as the variation of the constant of nature and the presence
of unobserved long range forces. Efforts to solve this difficulties 
by coupling the quintessence field to the electromagnetic
field \cite{carroll}, and to the curvature of the metric \cite{chiba}
have been made. Further, a time dependent but otherwise smooth scalar 
field such as those studied so far, are somehow unphysical as they
violate the principle of equivalence -the Q--matter must experience
clustering in some degree and so it cannot be entirely smooth.
Therefore to respect the equivalence principle one should 
assume that $\phi$ varies with position as well. Accordingly one should 
be led to forsake the FLRW metric an take up some inhomogeneous one 
instead, only that in such a case the computational effort is bound to 
be enormous and most likely no exact solution will emerge.

Despite that the Q--matter proposal cannot be regarded at this moment with
unreserved confidence, we feel this idea is still worth exploring
in the hope that the aforesaid difficulties may soon find a 
satisfactory answer.

\section*{Acknowledgments}
This work has ben patially supported by the Spanish Ministry of Education
under grant PB94-0718. LPC and ASJ thank the University of Buenos Aires
for partial support under project TX-93.

%==========================================================================


\begin{thebibliography}{99}
%==========================================================================

\bibitem{Perlmutter98}
S. Perlmutter, {\it et al.}, Bull. Am. Astron. Soc. (1997);
S. Perlmutter, {\it et al.}, Astrophys. J. {\bf 517}, 565 (1999).

\bibitem{Riess98} 
A.G. Riess, {\it et al.}, Astrophys. J. {\bf 116}, 1009 (1998).

\bibitem{Garnavich98} 
P.~M. Garnavich, {\it et al.}, Astrophys. J. {\bf 509}, 74 (1998).

\bibitem{OS95} 
J.~P. Ostriker and P.~J. Steinhardt, Nature {\bf 377}, 600
(1995).

\bibitem{Peeb84} J.P.E. Peebles, Astrophys. J. {\bf 284}, 439 (1984).

\bibitem{Wang99} 
L. Wang, R. Caldwell, J.P. Ostriker, and P.J. Steinhardt, Astrophys. J.
{\bf 530}, 17 (2000).

\bibitem{CDS98}
R.~R. Caldwell, R. Dave and P.~J. Steinhardt, Phys. Rev. Lett. 
{\bf 80}, 1582 (1998).

\bibitem{Perlmutter99}
S. Perlmutter, M.~S. Turner and M. White,
Phys. Rev. Lett. {\bf 83}, 670 (1999).

\bibitem{Scott}
S. Dodelson, M. Kaplinghat, and E. Stewart, ``Tracking oscillating 
energy", report astro-ph/0002360.

\bibitem{Bucher98}
M. Bucher and D. Spergel, Phys. Rev. D {\bf 60}, 043505 (1999).

\bibitem{Bucher99}
R. A. Batye, M. Bucher, and D. Spergel, ``Domain wall dominated universes",
report astro-ph/9908047.

\bibitem{Waga} I. Waga and P. M. R. Miceli,
Phys. Rev. D, {\bf 59}, 103507 (1999).

\bibitem{Peacock} J. A. Peacock, {\em Cosmological Physics} (Cambridge
University Press, Cambridge, 1999).

\bibitem{CMBdata}  
P. Steinhardt, in {\em Critical Problems in Physics},
eds. V.L. Fitch and D.R. Marlow (Princeton University Press,
Princeton, 1997).

\bibitem{zlatev}
I. Zlatev, L. Wang and P.J. Steinhardt,
Phys. Rev. Lett.  {\bf 82}, 896 (1999); 
I. Zlatev and P.J- Steinhardt, Phys. Lett. B {\bf 459}, 570 (1999);
J.P. Steinhardt, L. Wang and I. Zlatev, Phys. Rev. D {\bf 59},
123504 (1999).

\bibitem{dw93} D. Pav\'{o}n and W. Zimdahl, Phys. Letters A {\bf 179},
261 (1993).

\bibitem{self}
D.N. Spergel and P.J. Steinhard,
Phys. Rev. Lett. {\bf 84}, 3760 (2000);
J. P. Ostriker,
{\it ibid.} (to be published), astro-ph/9912548;
S. Hannestad, ``Galactic halos of self--interacting
dark matter", report astro-ph/9912558;
C. Firmani {\em et al.}, Mon. Not. R. Astr. Soc. (to be published),
astro-ph/0002376.

\bibitem{Moore}
B. Moore {\em et al.}, Astrophys. J. {\bf 535}, L21 (2000).

\bibitem{Barrow} J. D. Barrow, Nucl. Phys. B. {\bf 310}, 743 (1988);
D. Pav\'{o}n, J. Bafaluy and D. Jou, Class. Quantum Grav. {\bf 8}, 
347 (1991); L.P. Chimento, A.S. Jakubi and D. Pav\'on,
Class. Quantum Grav. {\bf 16}, 1625 (1999).

\bibitem{Ellis}
G.F.R. Ellis and J. Wainwright, in
{\it Dynamical Systems in Cosmology\/},
eds. J. Wainwright and G.F.R. Ellis
(Cambridge University Press, Cambridge, 1997).

\bibitem{Neta99}
N.A. Bahcall {\em et al.}, Science
{\bf 284}, 1481 (1999); J.R. Primack,
in {\em Cosmic Flows: Towards an Understanding of Large-Scale
Structure}, eds. S. Courteau, M.A. Strauss, and J.A. Willick
(ASP Conference Series, to appear), report astro--ph/9912089.


\bibitem{Efst}
G. Efstathiou, in {\em Structure formation in the Universe},
eds. N. Turok and R. Crittenden, Proceedings of NATO ASI (in the press),
report astro-ph/0002249.

\bibitem{zimdahl}
W. Zimdahl,
Phys. Rev. D, {\bf 61}, 083511 (2000).

\bibitem{d1}
W. Israel and W. Stewart, Ann. Phys. {\bf 118}, 341 (1979);
D. Pav\'{o}n, D. Jou  and J. Casas-V\'{a}zquez, Ann. Inst. H. 
Poincar\'{e}  A {\bf 36}, 79 (1982); V. Romano and D. Pav\'{o}n,
Phys. Rev.  D {\bf 50}, 2572 (1994).

\bibitem{m2}
R. Maartens,
in {\em Hanno Rund Workshop on Relativity and
Thermodynamics}, ed. S.D. Maharaj (Natal University, Durban, 1997).

\bibitem{mdv} 
M. Anile, D. Pav\'{o}n and V. Romano, 
``The case for hyperbolic theories of dissipation in relativistic
fluids", report gr-qc/9810014; D. Pav\'{o}n, in {\em Relativity and 
Gravitation in General}, eds. J. Mart\'{\i}n, E. Ruiz, F. Atrio and
A. Molina (World Scientific, Singapore, 1999).  

\bibitem{jpa}
D. Pav\'{o}n, D. Jou and J. Casas-V\'azquez, J. Phys. A 
{\bf 16}, 775 (1983).

\bibitem{HL}
W.A. Hiscock and L. Lindblom, Ann. Phys. {\bf 151}, 466 (1983);
{\em ibid.}, Contemporary Mathematics {\bf 71}, 181 (1988).

\bibitem{Stecker}
F.W. Stecker, D.L. Morgan, Jr. and J. Bredekamp,
Phys. Rev. Lett. {\bf 27}, 1469 (1971).

\bibitem{Chi93}
L.P. Chimento and A.S. Jakubi,
Class. Quantum Grav. {\bf 10}, 2047 (1993).

\bibitem{ChJMM}
L.P. Chimento and A.S. Jakubi, Class. Quantum Grav. {\bf 14}, 
1811 (1997); L.P. Chimento, A.S. Jakubi, and V. M\'{e}ndez,
Int. J. Mod. Phys. D {\bf 7}, 177 (1998); L.P. Chimento, A.S. Jakubi,
V. M\'{e}ndez, and R. Maartens, Class. Quantum Grav. {\bf 14}, 
3363 (1997).


\bibitem{textbook}
See any standard textbook on thermodynamics.

\bibitem{Bar93}
J.~D.~Barrow, Class. Quantum Grav. {\bf 10}, 279 (1993).

\bibitem{Chi96}
L.P. Chimento and A.S. Jakubi, 
Int. J. Mod. Phys. D {\bf 5}, 71 (1996).

\bibitem{RatraPeebles}
B. Ratra and P.J.E. Peebles, Phys. Rev. D {\bf 37}, 3406 (1988).

\bibitem{Guanajuato94}
L.P. Chimento and A.S. Jakubi, in 
{\em Recent Developments in Gravitation and Mathematical Physics},
eds. A. Mac\'{\i}as, T. Matos, O. Obreg\'on and H. Quevedo,
Proceedings of the First Mexican School on Gravitation and Mathematical
Physics (World Scientific, Singapore, 1996).

\bibitem{powerlaw}
M. Kaplinghat {\em et al.}, Phys. Rev. D {\bf 59}, 043514 (1999).

\bibitem{Cesari} 
L. Cesari, {\em Asymptotic Behavior and Stability Problems in Ordinary
Differential Equations} (Springer, Berlin, 1963). 

\bibitem{CHSN}
T. Chiba, N. Sugiyama and T. Nakamura,
Mon. Not. R. Astr. Soc. {\bf 301}, 72 (1998).

\bibitem{Ferreira}
P. G. Ferreira and M. Joyce, Phys. Rev. Lett.
{\bf 79}, 4740 (1997); C. Wetterich, Nucl. Phys. B {\bf 302},
668 (1988);  J. Copeland, A.R. Liddle,
and D. Wands, Phys. Rev. D {\bf 57},  4686 (1998).

\bibitem{carroll}
S.M. Carroll, Phys. Rev. Lett. {\bf 81}, 3067 (1998).

\bibitem{chiba}
T. Chiba, Phys. Rev. D {\bf 60}, 083508 (1999).

\newpage

\noindent
{\Large \bf Figure Captions}

\bigskip 
\noindent 
Figure 1.
The adiabatic index $\gamma_\phi$ of the quintessence scalar field {\em versus}
the matter density parameter $\Omega_m$, for CDM ($\gamma_m=1$), in the 
asymptotic attractor regime  for the model presented in Sec. IV A.

\bigskip 
\noindent 
Figure 2.
The adiabatic index $\gamma_\phi$ of the quintessence scalar field
{\em versus} the interaction rate parameter $\nu^{-1}$ and the 
dissipative contribution to the sound speed $v$, for CDM ($\gamma_m=1$),
in the asymptotic attractor regime, for the model presented in Sec. IV B.

\end{thebibliography}
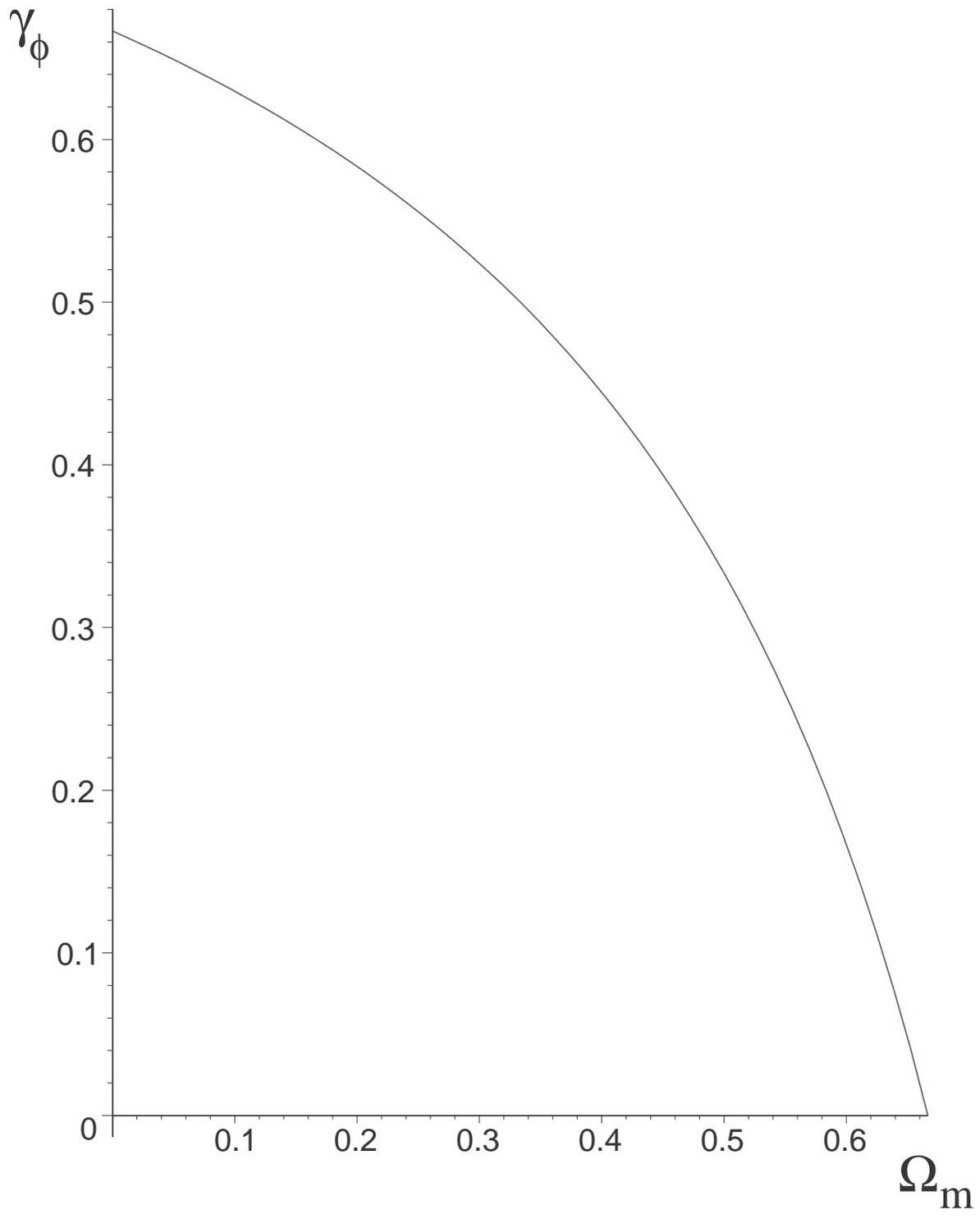
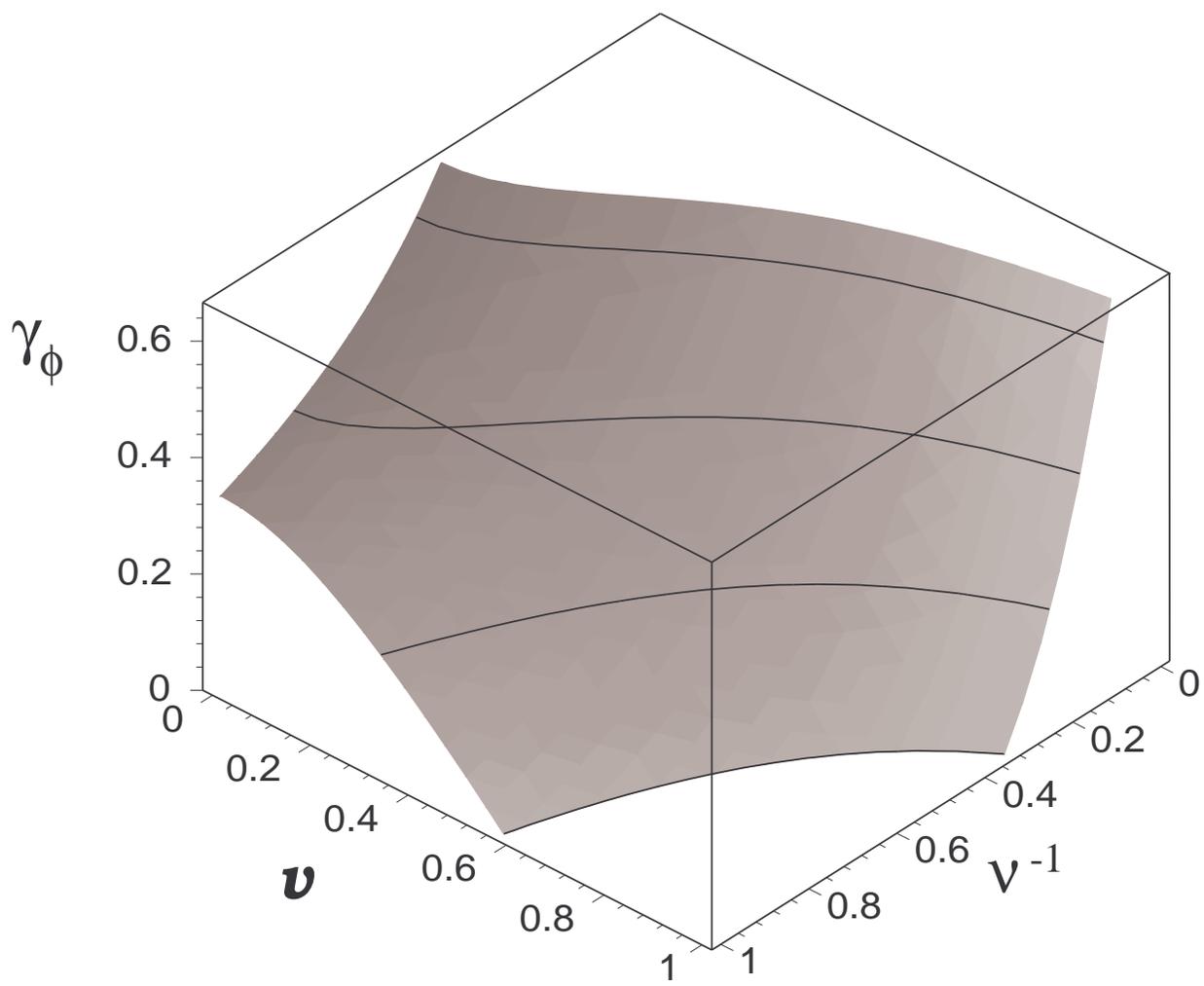
\end{document}